\documentclass{sig-alternate-2013}
\usepackage{caption}
\usepackage[skip=0pt]{subcaption}
\DeclareCaptionType{copyrightbox}
\usepackage{xspace}
\usepackage{hyperref}
\usepackage{paralist}
\usepackage{graphicx}
\usepackage{framed}
\usepackage{multirow}
\usepackage{url}
\usepackage{breakurl}
\usepackage{comment}
\usepackage{fancyvrb}

\long\def\com#1{}

\long\def\xxx#1{}

\long\def\abbr#1#2{#2}     

\renewcommand{\paragraph}[1]{%
  \vspace{0.5\baselineskip plus 0.7\baselineskip minus 0.2\baselineskip}%
  \noindent\textbf{#1.} }

\newcommand{\Name}{ConScript\xspace}
\newcommand{\name}{ConScript\xspace}

\abbr{
\newfont{\mycrnotice}{ptmr8t at 7pt}
\newfont{\myconfname}{ptmri8t at 7pt}
\let\crnotice\mycrnotice%
\let\confname\myconfname%
\permission{Permission to make digital or hard copies of part or all of this
work for personal or classroom use is granted without fee provided that
copies are not made or distributed for profit or commercial advantage, and
that copies bear this notice and the full citation on the first page.
Copyrights for third-party components of this work must be honored. For all
other uses, contact the owner/author(s). Copyright is held by the
author/owner(s).} 
\conferenceinfo{WPES'13,}{November 4, 2013, Berlin, Germany.} 
\copyrightetc{ACM \the\acmcopyr}
\crdata{978-1-4503-2485-4/13/11. \\
  http://dx.doi.org/10.1145/2517840.2517866}
}{}

\makeatletter
   \abbr{}{\def\@copyrightspace{\relax}}
\makeatother

\title{Conscript Your Friends into\\
  Larger Anonymity Sets with JavaScript}
\abbr{}{
\subtitle{(Extended Version)}
}

\numberofauthors{2}

\author{
  \alignauthor
	Henry Corrigan-Gibbs%
  \titlenote{Work conducted while author was a staff member at Yale University.}\\
  \affaddr{Stanford University}\\
  \email{henrycg@stanford.edu}
  \alignauthor
  Bryan Ford\\
	\affaddr{Yale University}\\
  \email{bryan.ford@yale.edu}
}

\begin{document}
\maketitle


\begin{abstract}
\xxx{Remove xxx marks before submitting}
\xxx{Cite extended version of paper}
\xxx{Get rid of widow section headers}
We present the design and prototype implementation of \name, a framework for
using JavaScript to allow casual Web users to participate in an 
anonymous communication system.
When a Web user visits a cooperative Web site, the
site serves a JavaScript application that instructs the
browser to create and submit ``dummy'' messages into the anonymity system.
Users who want to send non-dummy messages through the anonymity system
use a browser plug-in to replace these dummy messages with real messages. 
Creating such {\em conscripted anonymity sets} can increase the anonymity set
size available to users of remailer, e-voting, and verifiable shuffle-style
anonymity systems. 
We outline \name's architecture, we address a number of potential attacks
against \name, and we discuss the ethical issues related to deploying such a
system.
\com{
To demonstrate the compatibility of our framework with different underlying
anonymity systems, we have implemented a prototype \name client for a mix-net
and for a DC-net-style anonymity system.
}
Our implementation results demonstrate the practicality of \name: a
workstation running our prototype \name JavaScript client generates a dummy message for
a mix-net in 81 milliseconds and it generates a dummy message for a
DoS-resistant DC-net in 156 milliseconds.
\end{abstract}



\section{Introduction} 
Although anonymity systems 
based on verifiable shuffles~\cite{wolinsky12dissent} and
delayed message forwarding~\cite{danezis03mixminion}
may offer strong privacy guarantees,
the high end-to-end latency that these systems impose
makes them relatively unpopular.
Thus, users of Mixmaster~\cite{moeller00mixmaster},
Mixminion~\cite{danezis03mixminion}, 
and other such systems,
may enjoy strong anonymity, but only amongst a small number of users.
In many real-world situations, being
an anonymous within a small set of users is
almost as bad as having no anonymity at all,
especially since surveillance agencies may give extra
scrutiny to an anonymity system's encrypted traffic flows~\cite{nakashima13new}.
In contrast, low-latency anonymity systems 
such as Tor~\cite{dingledine04tor},
have relatively large user bases but
provide no protection against ISP-level adversaries~\cite{murdoch07sampled}.
A whistleblower trying to ``leak'' documents anonymously
is left to choose between 
unpopular anonymity systems with relatively strong security properties
and more popular systems which are vulnerable to low-cost traffic-analysis attacks.

To help increase the size of anonymity sets in 
strong anonymity systems---and thus
to make these systems more useful in practice---%
we propose forming {\em conscripted anonymity sets} 
using JavaScript.
Our framework, called \name, is compatible with a number of
anonymity systems, so we describe the high-level
ideas in the context of a generic anonymity system.
Later on, we discuss how to apply the generic
framework to popular anonymity systems.

Like AdLeaks~\cite{roth13secure}, an
independently developed architecture focusing on document leaking, \name{} 
leverages JavaScript programs served by 
Web servers to increase anonymity set size.  
In contrast with AdLeaks, \name is compatible with
existing anonymity systems, \name offers some protection against active attacks
by malicious insiders, and \name avoids the problem of message collisions and
the need for error-correcting codes in AdLeaks.

In \name, cooperative 
Web servers host a JavaScript 
application containing the anonymity system's client code. 
Whenever a user browses to a cooperating Web site, 
the JavaScript application instructs the user's {\em browser} to
function as a client of the anonymity system.
The browser performs the encryption and processing necessary to create
a ``dummy'' client message, then it submits the message to the
underlying anonymity system via the {\tt XMLHttpRequest} mechanism.
To avoid enlisting users against their will, the Web server may
obtain the explicit consent of Web users before running the 
\name JavaScript.

Actual users of the anonymity system, who want to send 
messages anonymously through the system, use a browser
plug-in to intercept the JavaScript client's dummy message 
and replace it with a real message for the anonymity system.
If these ``real'' messages are indistinguishable from the
conscripted user's ``dummy'' messages, the effective number
of participants in the anonymity system (from the perspective
of an adversary) is equal to the number of honest real users {\em plus}
the number of conscripted users.
Thus, actual users can hide amongst a 
much larger set of casual users browsing the Internet.
Careful construction of the plug-in protects against
arbitrarily malicious Web servers (who try to distinguish real users
from conscripted users), eavesdroppers, and adversarial clients.

When paired with a compatible anonymity system, using \name{}
{\em can only increase} the anonymity given to a particular client.
In the worst case, the anonymity that a client gets is equal
to the total number of real users---i.e., {\em no worse}
than it would have been without using \name.

To demonstrate the practicality of \name, we have implemented proof-of-concept
JavaScript clients for {\em two different} underlying anonymity systems:
a mix-net and a DoS-resistant client/server DC-net~\cite{corrigangibbs13verif}.
These proof-of-concept applications are not wire-compatible
with the corresponding deployed systems---we use them only
to approximate the performance of the client application
in a deployed system.

\abbr{}{
We have tested \name's JavaScript 
application on a Linux workstation, a Mac laptop, 
an Apple iPhone, and an Android phone using
the Chrome, Firefox, Safari, and Opera Web browsers 
(where available).
On a Linux workstation using the Chrome Web browser, 
generating a mix-net dummy message 
takes 81~ms and generating a dummy message
for the DC-net-style system takes 156~ms.
On an iPhone running the Chrome Web browser, 
generating the mix-net dummy message takes
9,009~ms and generating the DC-net-style message
takes 62,973~ms.
We also compare the power required for generating dummy
messages to the power required for normal Web browsing activity.
On an Android phone, generating a DC-net-style dummy message
consumes $2.3\times$ more energy than the energy consumed
while browsing a Webmail client for the same amount of time,
and message generation consumes $4.9\times$ more energy than is consumed
while browsing a news Web site.

This paper's contributions are:
\begin{compactitem}
  \item a general framework for using Web users to increase
        the anonymity set size of a number of existing anonymity systems,
  \item an analysis of a variety of attacks against conscripted
        anonymity systems and prevention techniques,
  \item discussion of the ethical issues surrounding conscripted
        anonymity,
  \item a prototype implementation of \name's JavaScript
        client for two different anonymity systems:
        a mix-net and a DoS-resistant client/server DC-net, and
  \item evaluation of \name on a variety
        of devices.
\end{compactitem}
}

Section~\ref{sec:arch} outlines \name's architecture.
Section~\ref{sec:attack} describes how \name defends against 
a number of possible attacks.
Section~\ref{sec:ethics} addresses ethical issues
related to conscripted anonymity,
Section~\ref{sec:impl} summarizes the results of
our implementation and evaluation, 
Section~\ref{sec:rel} describes related work, 
and Section~\ref{sec:concl} concludes.

\section{Architecture}
\label{sec:arch}

\abbr{}{
This section introduces \name's system participants, summarizes the 
trust assumptions we make, and discusses the compatibility
of \name with a number of pre-existing anonymity systems.

The overall goal of \name is to increase the anonymity
set size provided to the users of an underlying anonymity system
by conscripting casual Web users into participating in the
anonymity system.
\Name should provide these properties even if a number of
the participants collude to undermine
the anonymity of the system. 
}

\subsection{Participants}
\label{sec:arch:parts}

\begin{figure}
\centering
\includegraphics[width=0.5\textwidth]{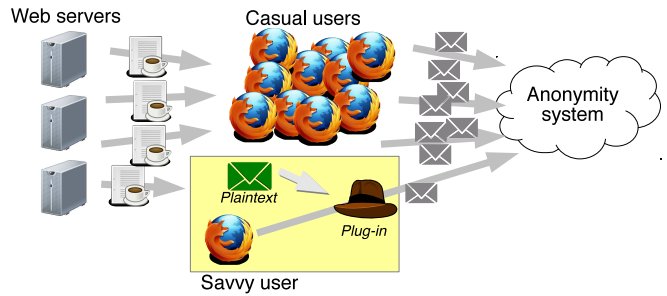}
\caption{Overview of \name's system architecture.}
\label{fig:model}
\end{figure}

\Name's architecture 
consists of three components:
an {\em underlying anonymity system},
a number of {\em cooperative Web servers},
and {\em Web users}.
Figure~\ref{fig:model} provides a pictorial
representation of the interaction of the
system's participants.

\paragraph{Anonymity system nodes}
At the core of \name's architecture is a
pre-existing anonymous messaging system.
\Name is, in principle, compatible
with a number of existing anonymity systems, but
to make the system design concrete, we will first
describe how to use \name
with a mix-net consisting of {\em single} cascade of
timed mixes~\cite{chaum81untraceable}. 

A mix cascade consists of a set of $M$ dedicated mix servers,
with each server $i$ having a well-known public key $pk_i$.
Over the course of a day (or some other time period), 
each of $N$ users submits a fixed-length 
message to a {\em message pool} hosted by the first mix-server.
Users serially encrypt each
message $m$ with each of the servers' $M$ public keys:
\begin{align*}
E(pk_1, \dots E(pk_M, m) \dots )
\end{align*}
At the end of the day, the first server shuffles the set of
ciphertexts in the message pool, removes duplicate messages,
decrypts a layer of encryption, and forwards
the messages to the second server.
This process continues until the last server holds
the $N$ plaintext messages in permuted order.
At this point, the last server could, for example,
post the anonymized messages on a public bulletin board. 
If at least one server does {\em not} collude with 
the others, then an honest sender's message is anonymous amongst
the $h \leq N$ honest senders
(provided that we ignore the possibility of active attacks 
by the mix servers).

An important property of a timed cascade mix-net, and similar systems,
is that the anonymity of a single sender 
{\em increases monotonically} with the number of
other senders who have submitted a message to the system.
That is, the anonymity set size of a given message in the 
{\em timed} mix-net system {\em never decreases}
when an additional sender adds a message to the message pool.
\abbr{}{
In certain cases (e.g., if the senders are 
Sybil identities~\cite{douceur02sybil}), 
adding messages to the pool does not increase the senders' 
anonymity set size,
but these new messages do not decrease their anonymity 
set size either.
}

\Name takes advantage of the monotonicity of the anonymity in
mix-net-style systems. 
Since adding more users to an anonymity system cannot decrease
a user's anonymity, it does not hurt
to gather messages from conscripted and unauthenticated 
senders and insert them into the anonymity system.

\paragraph{Web servers}
\Name requires the cooperation of 
a number of Web servers, each of which 
serves a \name JavaScript application to its Web clients
alongside the Web content it would normally serve.
Individuals and organizations interested in supporting
Internet privacy or in protecting whistleblowers
might agree to embed the \name JavaScript application 
in their Web sites.
For example, the Electronic Frontier Foundation,
Wikipedia, and the Guardian newspaper might each agree to 
serve the \name client script to users visiting
their Web sites.

The JavaScript application, embedded in a normal
Web page, contains the client code for
the anonymity system in use.
For example, if the anonymity system were a mix-net, the
JavaScript code would contain everything that a user
would need to generate a {\em dummy} message for the mix-net.

In the case of a mix-net, the JavaScript would contain
the public keys of the $M$ servers, a method to generate
a dummy plaintext message of the correct length, 
and the public-key encryption routines for encrypting the
plaintext message with each of the $M$ public keys.

All clients' {\em plaintext} messages in the mix system must 
begin with a single control bit indicating whether the 
message is a dummy message (\texttt{control\_bit=0})
or a real message (\texttt{control\_bit=1}).
In this way, the recipient of a message can determine
whether to discard the message as a dummy or process
it as a real plaintext.
The client-side \name JavaScript application 
encrypts the control bit alongside the clients' plaintext message. 
(We describe how users send non-dummy messages into the anonymity
system later in this section.)

We assume that the mix-net uses an IND-CCA2-secure public key
encryption scheme~\cite{rackoff91noninteractive} to 
ensure that the encrypted
ciphertexts leak no information about the plaintext
even under an adaptive chosen ciphertext attack.
In particular, the encryption of a dummy message 
(a string of zeros) and the encryption of a real message
(an arbitrary string beginning with a control bit set to one) 
must be indistinguishable under an adaptive chosen ciphertext attack.
A number of standard cryptosystems, including
RSA-OAEP, provide IND-CCA2 security under standard
hardness assumptions~\cite{fujisaki01rsa}.
All IND-CCA2-secure cryptosystems 
use {\em randomized} encryption routines, 
so the ciphertexts generated by encrypting two 
dummy plaintexts (strings of zeros)
will be different with overwhelming probability.

The JavaScript application also contains the
code to submit the final message to the
first mix server using the {\tt XMLHttpRequest} API.
This HTTP request will be a {\em cross-origin
request}, since the script issuing the {\tt GET} request
was served by the cooperative {\em Web server}, but the target
of the {\tt GET} request is the {\em mix server}.
The ``same-origin policy'' would normally prevent the 
Web browser from communicating directly with the
mix servers, but the mix nodes
can include the {\tt Access-Control-Allow-Origin} 
header in their HTTP response to allow 
this sort of cross-origin request~\cite{mozilla-cors,www13cross}.
\abbr{}{
The process of submitting a client message to the mix servers 
will proceed as follows:
\begin{compactenum}
  \item The cooperative Web server at {\tt web-server.net} 
        serves the \name JavaScript application to 
        the client's browser.
  \item The JavaScript application creates a dummy message at
        the client.
  \item The client issues a cross-origin {\tt XMLHttpRequest} to 
        {\tt mix-server.net} containing the dummy message.
  \item The client's browser issues a ``pre-flight'' request
        to {\tt mix-server.net} to ensure that the mix server
        accepts cross-origin requests (see \cite[Section 7.1.5]{www13cross}).
  \item The mix server returns an {\tt Access-Control-Allow-Origin: *}
        header to the client, along with a status code (e.g, {\tt 200 OK}).
  \item The client's browser receives the {\tt Access-Control-Allow-Origin}
        header and passes the cross-origin response data to the 
        client's JavaScript.
  \item The mix server at {\tt mix-server.net} adds the client's
        message to its input pool.
\end{compactenum}
}

\paragraph{Casual users}
Casual users, the first class of \name users, 
are normal Internet users using
standard JavaScript-enabled Web browsers with 
no special browser extensions or modifications.

When a casual user visits the target Web page
(hosted by a cooperative Web server as described above),
the user's browser will download the page content,
which includes the \name client JavaScript application.
Once the casual user's browser downloads the \name client
application, the script will
cause the browser to send mix-net-encrypted dummy messages into the anonymity
system. 
In practice, the script might ask for the user's permission
before starting to send messages into the anonymity system
(see Section~\ref{sec:ethics}).
After the script begins running, the
casual client will not notice any out-of-the-ordinary
behavior, except for perhaps a slight drop in browser
performance due to the computational burden of generating
the mix-net messages.

In this way, the casual user becomes a 
client of the underlying anonymity 
system without needing to download a browser
extension or install any software tools.
Casual users need not know how to generate a public key,
install a program, or configure anonymity system client software.
Since the casual users' browsers submit genuine client
messages into the anonymity system, these casual
users are 
(with caveats explained in Section~\ref{sec:attack}) 
indistinguishable from real users, from the
perspective of an eavesdropping adversary.

\paragraph{Savvy users}
Savvy users, the second class of \name users,
send non-empty messages through the anonymity system,
as opposed to the dummy messages that
the casual users send.
The only difference between a casual user
and a savvy user is that every savvy user has
a plug-in installed in their Web browser that
monitors the browser's outgoing HTTP requests.
To send a message through the anonymity system,
the savvy user enters their secret message into the plug-in
and then browses to a cooperating Web site.
When a savvy user visits a Web page that contains
our system's JavaScript client, the plug-in will
transparently {\em intercept} the outgoing 
message from the browser and
will replace the dummy message (generated by
the JavaScript client) 
with a content-carrying message
generated by the plug-in.

The effectiveness of \name relies on the difficulty of distinguishing
savvy users from casual users, so we take a number
of steps to prevent side-channel and equivocation
attacks that would allow a malicious Web server or
an eavesdropper to identify which users have
the plug-in installed.
Section~\ref{sec:attack} describes these defenses.

\subsection{Trust Assumptions}
\abbr{}{
One attractive feature of \name's architecture is that
it requires participants to make only 
minimal trust assumptions beyond those required
by the underlying anonymity system.
We classify our assumptions into two categories:
the assumptions required to ensure that using \name{}
{\em does not reduce} the anonymity provided to a system
user, and the assumptions required to guarantee that
\name{} {\em increases} the anonymity set size of a system.
}

We say that a participant is ``honest'' if it executes operations
as the system design dictates and if it does not collude with other
nodes or an external adversary.

To guarantee that using \name does not reduce the
level of anonymity provided by an underlying system,
we must assume that honest savvy \name users can access
{\em at least one} honest \name-enabled Web server.
This Web server must, in turn, be able to
communicate with the underlying anonymity system.
If savvy clients cannot access any cooperative Web server,
then these clients cannot submit their messages to the
underlying anonymity system, and the anonymity set size
provided will be smaller than the total number of savvy users.

To guarantee that using \name has the potential to 
{\em increase} the anonymity set size of the underlying
anonymity system, we must additionally assume that
at least one honest casual client must be able to
communicate with at least one honest Web server. 
This honest Web server must be able to 
communicate with the underlying anonymity system.

\abbr{
We make no additional assumptions about user or 
server behavior, though 
the underlying anonymity system might require additional
trust assumptions over and above those we must make.
For example, a mix-net cascade requires that
at least one of the mix servers is honest.
}{
We make no additional assumptions about user or 
server behavior.
Users (both casual and savvy) in our model can be 
{\em arbitrarily malicious}---they can 
submit malformed messages, they can collude
to submit many copies of the same message,
and they can cooperate with the Web server
to try to undermine the anonymity of some
other user.

Participating Web servers can also be 
arbitrarily malicious---they can serve
incorrect JavaScript to users, they can 
serve different JavaScript to different users,
and they can selectively deny service 
to some users (we discuss this class of attack
in Section~\ref{sec:attack:dos}).
To maintain the security of the underlying
anonymity system, we only require, as stated above, that 
savvy clients can communicate with at least one
honest Web server.

The underlying anonymity system might require additional
trust assumptions over and above those we must make.
For example, a mix-net cascade requires that
at least one of the mix servers is honest.
}

\subsection{Underlying Anonymity System}
\Name is compatible with any anonymity system 
that has a certain set of properties, which
we enumerate below.

\paragraph{Anonymity set size is monotonic w.r.t. users}
The anonymity provided to
a particular user of the anonymity 
system must {\em increase monotonically}
with the number of total users of the system.
If the anonymity of a particular user can {\em decrease}
when the system has more (potentially adversarial) users, then conscripting
many users into the anonymity system might
actually {\em hurt} the anonymity of the system overall.

\paragraph{Simulatable traffic streams}
It must be possible to simulate the behavior of a
real user such that the behavior of the 
simulated user and the real user are 
indistinguishable from the perspective of an adversary.
For example, the mix-net client simulator generates an
onion encryption of a string of zeros using an IND-CCA2-secure
cryptosystem.

\paragraph{Easy to identify malformed messages}
The anonymity system should be able to identify and
reject malformed user messages (to prevent
a malicious user from disrupting communication).

\paragraph{Messages do not depend on the set of active users}
To submit a message to the anonymity system,
a user should not need to know the identities
of the system's other users.
A traditional DC-net~\cite{chaum88dining}, for example, would not be
suitable because it requires every user to share
a secret with every other user of the system.
In contrast, the client/server DC-net we use
only requires users to know the public keys of the system's 
servers~\cite{corrigangibbs13verif}.

\subsection{Compatible and Incompatible Anonymity Systems}

We now briefly describe which 
anonymity systems are compatible with \name.

\paragraph{Yes: Timed Cascade Mixes and Verifiable Shuffles}
Timed cascade mixes (introduced in Section~\ref{sec:arch:parts}) 
and verifiable shuffles~\cite{brickell06efficient,neff01verifiable}
satisfy all properties necessary to be compatible with \name.

\paragraph{Probably Yes: Anonymous Remailers}
Any anonymous remailer using fixed-length messages which has the monotonic
anonymity property described above is compatible with \name.
Not all anonymous remailers exhibit the monotonic 
anonymity property, however, and
determining whether or not a particular remailer system has this property 
is not necessarily straightforward in general.
\abbr{}{
Under a more restrictive threat model---in which
the adversary controls a some constant fraction of nodes in the network, 
for example---%
it might be safe to assume that a remailer system
satisfies the monotonic anonymity property.
We do not claim that any particular remailer has the monotonic
anonymity property, but we assert that {\em if} any remailer system does
exhibit the property, then it would be suitable for use with \name.

}
To create a dummy message for an anonymous remailer,
the client JavaScript application follows a process
very similar to the process used to create a dummy
message for the mix net (described earlier in this section).
\abbr{}{
The dummy plaintext is simply a string of zero bytes, padded
up to the fixed length of a remailer plaintext.
The client prepends a control bit to its message, indicating
whether or not the message is a dummy.
The client then iteratively encrypts this message with the 
public keys of each remailer server, using an IND-CCA2-secure cryptosystem. 
By the indistinguishability property of the cryptosystem,
an encrypted dummy message and a real message will be indistinguishable.
}
We discuss other design issues related to using
remailers in Section~\ref{sec:impl}.

\paragraph{Probably No: Tor}
\abbr{
It is not clear whether any anonymity gain would result
from conscripting users into the Tor network.
}{
To be useful with \name, there
must be an efficient mechanism for a casual client to generate
a dummy traffic stream into the anonymity system that is indistinguishable
from the stream that a savvy user would generate.
While generating dummy traffic in mix-net-style systems is straightforward,
it is less clear what the analogous dummy traffic stream would look like 
for Tor~\cite{dingledine04tor}.
A ``straw-man'' traffic simulator might try to simulate HTTP browsing
behavior by downloading a number of popular Web sites through the
Tor tunnel in random order.
However, an adversary could likely use packet timing information
to distinguish these simulated HTTP traffic streams from real streams,
which might carry a wider variety of traffic types.

In addition, it is not clear whether any anonymity gain would result
from conscripting users into the Tor network.
}
If a Tor user picks a route through the Tor network with an 
adversarial first and last node, that user's anonymity is compromised.
If some fraction of the relays in the Tor network are dishonest, 
adding more clients to the Tor network {\em does not} change
the probability that a new circuit selected through the
network will begin and end at an adversary-controlled 
relay~\cite{syverson09entropist}. 
Conscripting users into the Tor network might actually {\em harm}
real users, since the conscripted users would consume Tor's scarce
network resources with dummy messages, leaving less network bandwidth for
real users' messages.

\abbr{}{
\paragraph{No: Threshold Mixes}
In a threshold cascade mix, the first mix server collects $n$ messages in
its input pool before mixing the messages, removing a layer of
encryption, and sending the messages on to the next mix 
server~\cite{serjantov03trickle}.
Threshold mixes are {\em not} suitable for use with \name
because the anonymity of the mix is not monotonic in the number of users.
If there are $n$ honest users of the mix, the anonymity set size 
provided to each sender is $n$.
If one dishonest user is able to insert a message into the mix before
the honest users begin submitting messages to the mix, then the mix
will fire after $n-1$ {\em honest} users have submitted messages to the mix.
The effective anonymity set size for these users then decreases to 
$n-1$ from $n$.
Adding a dishonest user to the system {\em decreases} the anonymity set size
provided to existing users of the system.
Conscripting many potentially dishonest users into a threshold mix system
could weaken the anonymity of the users of that system.
}

\subsection{Effective Anonymity Set Size}

\abbr{}{
As long as the behavior of a casual
user is {\em indistinguishable}
from the behavior of a savvy user---%
{\em even if} the Web server and many
other users are malicious---%
the savvy users can ``hide amongst'' the casual users.
}

Consider a deployment of \name that has $j$ (honest) savvy users and 
$k$ (honest) casual users,
whose messages reach the anonymity system servers.
\Name provides a level of anonymity 
equivalent to the level that would be offered 
by the underlying system when run with $j$ real users
plus $k$ users who submit ``dummy'' messages into the system.
In the simple cascading mix-net we described earlier,
the effective anonymity set size would be $j+k$.

\abbr{}{
\subsection{Limiting Participation Rate}
\label{sec:arch:rate}
The \name JavaScript client runs whenever a casual
user visits a \name-enabled Web site and has consented
(explicitly or implicitly) to participation in the system.
Although the \name client could instruct the casual 
client's browser to submit messages into the underlying anonymity
system as often as possible, this ``always on'' policy is just
one of many possible participation policies.
The \name client could instruct the browser to generate a
message with probability $1/t$, so that the browser generates
a message after a $t$ visits to the Web site, in expectation.
This rate-control technique would also enable the Web server
to the rate at which users (both casual and savvy) send messages
into the underlying anonymity system, which could prevent a flood
of honest casual users from overwhelming the system.

Since battery life is a concern on mobile devices, the 
Web server could also impose more severe rate limits on mobile devices
to prevent the \name JavaScript from quickly consuming
a casual mobile user's battery.
(We evaluate \name's impact on battery life in Section~\ref{sec:impl}.)
}

\section{Attacks and Defenses}
\label{sec:attack}

In this section, we consider possible attacks against
\name, leaving attacks against the underlying anonymity system out of scope.

\subsection{Malformed JavaScript}
\label{sec:attack:javascript}

An adversarial Web server might modify the 
JavaScript it sends to the client in an attempt
to distinguish savvy users (who have the conscripted
anonymity set browser plug-in installed) from casual
users (who do not have the plug-in).
\abbr{The extended version of this paper shows 
a sample of malicious JavaScript that a malicious
Web server might send to try to distinguish savvy users
from casual users.}{For example, 
the pseudo-code in Figure~\ref{fig:code-correct} 
shows the correct JavaScript for generating dummy messages.
Figure~\ref{fig:code-evil} shows the code that a malicious
Web server might send to try to distinguish savvy users
from casual users.

\begin{figure}[t]
\centering
\begin{subfigure}{0.45\textwidth}
\begin{Verbatim}[frame=single,framesep=5pt]
function MakeAnonymityMessage() {
  // Code to generate a dummy message
  // ...
  return dummy_msg;
}

// Plug-in replaces MakeAnonymityMessage()
var msg = MakeAnonymityMessage();
var success = UploadToServer(msg);
\end{Verbatim}
\caption{Correct code}
\label{fig:code-correct}

\end{subfigure}

\vspace{1em}

\begin{subfigure}{0.45\textwidth}
\begin{Verbatim}[frame=single,framesep=5pt]
function MakeAnonymityMessage() {
  return "Bogus!";
}

// Plug-in replaces MakeAnonymityMessage()
var msg = MakeAnonymityMessage();

// Distinguishing attack
if(msg == "Bogus!") {
  // User is a casual user
} else {
  // User is a savvy user (has plug-in)
}

var success = UploadToServer(msg);
\end{Verbatim}
\caption{Evil code}
\label{fig:code-evil}
\end{subfigure}

\caption{Examples of correct and evil JavaScript code
served by a Web server in \name.}
\end{figure}

}

To defeat these attacks, the browser
plug-in must {\em only} replace
the dummy message with the savvy user's real message 
when the JavaScript on the relevant page 
{\em exactly matches} the JavaScript that the plug-in
expects to see.
To perform this check, the browser plug-in must have a copy of
the JavaScript code that it expects the Web server to send.
If the Web server sends JavaScript code
that does not match the expected code, the plug-in should simply
run the JavaScript as a casual user would.

The requirement that the plug-in have a copy of the
Web server's expected code is somewhat burdensome---it
restricts the type of content that the Web server can
serve alongside the \name JavaScript application.
A modern Web page often has tens of linked scripts,
{\tt iframe}s, Flash movies, and other dynamic content, but
a malicious Web server could exploit any of these objects
running alongside the JavaScript
client to mount a distinguishing attack.
Preventing the distinguishing attack requires the plug-in
to have a copy of {\em all} content on the page
(except for static text and images).
In this way, the plug-in can detect when the server
has served malicious or incorrect JavaScript to the user.

\abbr{}{
The static text and images on the site do not 
need to be bundled with the plug-in because a Web
server that serves different static content to different
Web users gains no advantage in distinguishing savvy
users from casual users.
If the casual and savvy users both are running the
same correct \name JavaScript code but the server sends different
static images to different clients, a casual client
will handle the static content in exactly the same way 
that a savvy client would, since the \name client
does not read or execute static page content.

In the future, there may be a way to use the
{\tt iframe}'s {\tt sandbox} attribute in HTML5
to allow the Web server to serve rich content along
with the JavaScript client.
For example, the JavaScript client could run in 
the parent frame and the server's rich content could
run inside of a {\tt sandbox}ed {\tt iframe}.
In this way, the server could provide interactive content
while also supporting conscript anonymity sets.
We leave a rigorous treatment of 
this application of sandboxing for future work.
}

\subsection{Selective Denial of Service}
\label{sec:attack:dos}
A malicious Web server could try to distinguish
savvy users from casual users by selectively 
denying service to users of to the anonymity system,
in an attack analogous to 
the {\em trickle attack} against mix-nets~\cite{serjantov03trickle}.
For example, if the attacker wants to determine
if a particular user is a savvy user, the Web server
could serve incorrect JavaScript to all users
except a particular target user and a
set of users that the attacker controls.

If the anonymity system outputs a real message when 
fed messages from only the target user as input, then the 
malicious Web server both learns that the target
user is a savvy user {\em and} the attacker learns the
content of the savvy user's message.

One technique to prevent such attacks is to maximize
the number of cooperating Web servers serving 
the JavaScript application.
Every savvy client could visit a number
of Web servers (instead of just one)
to ensure that a single malicious
Web server cannot block 
communication between users and 
the anonymity system.
Other techniques for preventing the trickle
attack could also apply here~\cite{serjantov03trickle}.

\subsection{Dangers of Cryptography in the Browser}
The application that generates casual 
clients' dummy messages must
implement public-key cryptography 
algorithms in JavaScript.
(In contrast, the savvy clients' messages 
are constructed by the browser plug-in,
which presumably can access standard
cryptography libraries.)

The Web browser environment is not the ideal place
to run cryptographic software:
many browsers do not offer a source of 
cryptographic randomness, it is difficult to prevent
side-channel attacks in the browser---%
perhaps mounted by a script running in another tab---%
and client-side cryptography libraries 
are less mature than their
server-side counterparts.
Even so, these limitations may not be fatal.
In the worst case, a flaw in or a side-channel
attack against the client-side cryptography library will
allow an adversary to distinguish the savvy
from the casual clients but such an attack
will {\em not} allow the adversary to read
the savvy clients' messages or to otherwise violate
the anonymity of the underlying system.

Since the savvy clients' messages are generated
using the browser plug-in, and since modern 
Web browsers allow plug-ins to execute native code 
through the Netscape Plugin Application Programming Interface (NPAPI), 
the savvy clients' messages will be encrypted using cryptographic
routines provided by conventional cryptography libraries (e.g., GPG).
Thus, in the worst case of an adversary who can
distinguish all casual clients from savvy clients, 
a savvy client will be still be anonymous amongst
the set of savvy clients, all of whom use the 
plug-in to encrypted their messages.

\abbr{
\subsection{Other Attacks}
The extended version of this
paper describes how a \name deployment
could defend against a number of
other possible attacks.
}{
\subsection{Sybil Attack}
\label{sec:attack:sybil}
Many anonymity systems will not release plaintext
messages until the estimated level of anonymity passes a certain threshold.
For example, a ``threshold and timed mix''~\cite{serjantov03trickle}
collects messages for at least $t$ seconds {\em and} until 
there are at least $n$ messages in the input pool before
processing the messages in the mix's pool.
If each of the $n$ messages comes from a different honest
user (and if we ignore other possible attacks), the 
size of each user's anonymity set is $n$.
However, if a single Sybil attacker~\cite{douceur02sybil}
can send $n-1$ messages into the mix 
(a {\em flood} attack~\cite{serjantov03trickle}), then 
the one honest user's effective anonymity set size will be one---%
the system provides no anonymity at all.

Verifiable shuffle~\cite{brickell06efficient} 
and e-voting schemes prevent these attacks by keeping
a ``roster'' of {\em registered users} public keys
and requiring a certain number of authenticated 
registered users to submit messages to the system
before processing the messages.

We can adopt a similar technique to prevent Sybil attacks.
The anonymity system maintains a roster of registered
users of the system.
Registered users (a subset of the savvy users) 
store their keypair in the conscripted anonymity set plug-in.
When a registered user enters their message into the
anonymity set plug-in, the plug-in presents them with
the offer to sign their message to the anonymity system.
The anonymity system could then collect $n$ messages
{\em signed} by registered users, plus any number
of messages from unregistered users, before it fires
the mix (or executes the verifiable shuffle operation).
To prevent the Web server and anonymity system from learning
which registered users submitted messages in a given
time period, registered users could authenticate using
a ``use-$l$-times'' pseudonym system~\cite{lysyanskaya00pseudonym}.

An important point is that registered users lose their
ability to hide amongst the casual users: since
registered users sign their messages to the anonymity
system, the Web server can distinguish registered users
by observing the signature, or lack of a signature, on a message.
Unregistered savvy users can still submit messages
to the system and hide amongst the casual users, it is
just that messages sent by savvy users do not count towards
the $n$-message threshold for firing the mix.

\subsection{Output Message Formatting}
If the \name JavaScript application and the \name
browser plug-in generate messages for the underlying 
anonymity system that are syntactically different, 
an attacker could use these syntactic
differences to mount a distinguishing attack.
For example, a mix-net ciphertext created by the 
client-side JavaScript application and
\name plug-in might be encoded as a JSON object.
If the JavaScript application and plug-in produce
JSON objects which use slightly different formatting
conventions (tabs versus spaces, single versus
double quotes, ordering of keys in a dictionary data structure,
etc.), an eavesdropping adversary
could use these formatting differences to distinguish
savvy users from casual users.
These subtle formatting differences could arise
if the JavaScript and plug-in applications use different
serialization or cryptography libraries, or they could
arise from implementation bugs.
To prevent this class of security weaknesses, the 
implementer of a \name deployment should ensure that there
is a single canonical form of \name output messages and
should make certain the JavaScript and plug-in applications
both generate messages that conform to that convention.

\subsection{Side-Channel Attacks}
There are a number of possible side-channels that
an adversary could use to distinguish casual
users from savvy users.
For example, casual users generate one
message when they load the client JavaScript
(a dummy message), while savvy users generate
two messages (a dummy message and then a real message).
If it takes twice as long for savvy users 
to generate their messages, an adversary could use
timing information to distinguish the casual
users from savvy users.

To prevent this particular attack, savvy clients
could pre-generate the real message that they will send
to the anonymity system servers
so that the time it takes for them
to run the client JavaScript and send their message
back to the Web server is nearly equal to the time
taken by a casual client.
This is just one of many side-channel attacks
that an implementation of \name must consider---we draw attention to
this class of attacks without attempting to address them all.

\subsection{Number of Savvy Users Leaked}

The casual users in \name submit ``dummy'' messages
into the system.
In a mix-net, a dummy message might
be a string of zeros, serially encrypted with the public keys of each
of the mix servers.
Once the mix servers process the $n$ messages in their input
pools, the last mix server will be able to count how many 
dummy messages there were in the input pool.
In this way, the last mix server can learn how many savvy 
users sent messages into the system.

The mix server might be able to use this information
to mount an {\em intersection attack}~\cite{kedogan02limits}
over a period of time to make an educated guess about
which users are savvy users and which are casual users.
Intersection attacks are a problem for every class of
anonymity system, so we leave protection against this 
class of attacks out of scope.

\subsection{Plug-in Distribution}
Since only savvy users will download the browser plug-in,
if the adversary can find out who has downloaded the \name
browser plug-in, the adversary can distinguish the savvy users
(who have the plug-in) from the casual users (who do not).
One possible way to prevent this attack would be to distribute
the plug-in using networks that the adversary cannot monitor.
For example, a group of acquaintances could share the plug-in software
by exchanging USB flash drives.
}

\section{Ethical Issues}
\label{sec:ethics}

Up to this point, we have considered only the technical
questions related to the deployment of \name,
but we have not addressed the 
equally important ethical issues that deployment
of such a system would raise.
The fundamental question is whether it is ethical
to ``conscript'' an unsuspecting Web user into 
participating in an anonymity system without the
user's consent.
Instead of trying to resolve this ethical question here,
we will outline three possible deployment scenarios of
\name (with varying levels of ``conscription'') and will 
make an ethical argument for each.

\paragraph{User opt out}
One possible way to deploy \name would be
to require the cooperating Web server to display a
conscripted anonymity ``badge'' on any Web page that serves
the \name JavaScript client.
Web users could ``opt out'' of participation in the
\name by clicking the badge.
FlashProxies~\cite{fifield12evading}, a system for using
Web browsers for censorship circumvention, takes this approach.

\abbr{}{
Another possibility would be to modify the Web site's ``terms
of service'' to indicate that by visiting the Web site, the user
implicitly consents to being conscripted into an anonymity system.
In this way, being conscripted into an anonymity system would constitute
a Web user's ``payment'' for visiting the Web site, much as Web users
today download ads and JavaScript trackers as payment for viewing
commercial Web content.
}

A utilitarian argument in favor of an ``opt-out'' approach is that the total
social benefit of \name is much greater than the total social cost
of conscription for the unsuspecting Web users.
This argument would be most persuasive in areas where the probable risk to a
conscripted Web user is low but where the social benefit of anonymous
communication is high.
For example, in a country with a judicial system 
that would not imprison a conscripted Web user just for being conscripted,
and with an invasive Internet surveillance regime, an ``opt-out'' policy might
be the most ethical one.

\abbr{}{
Another argument in favor of the opt-out deployment strategy is that using
opt-out might actually protect conscripted {\em and} savvy Web users by making
it easier to deny (e.g., to the secret police) that these users ever agreed to
use an anonymity system.
If users are required to opt in to \name, then this act of consent
might actually make the user legally liable for participating
in the anonymity system, 
which would make the user {\em more} vulnerable than if there had been no 
opt-in mechanism at all.
}

\paragraph{User opt in}
Another possible deployment strategy would be to require the explicit consent
of Web users before conscripting their browsers into the anonymity system.
For example, a pop-up window appearing after the Web page 
loads could explain the anonymity system to the user,
including the potential effects on the user's bandwidth and power usage,
and then ask whether the user wants to participate.

An ethical argument in favor of an ``opt-in'' deployment strategy is that a Web
user should have the choice of whether or not to participate in \name,
especially if participation could consume the user's bandwidth or drain
the battery on the user's mobile device.
Giving the user a choice to opt in to the network allows those users
who want to participate the option to do so, but does not 
force participation in an anonymity system on those who do not.
\abbr{}{This argument would be especially persuasive if the cost of being 
conscripted into an anonymity system is high.}
If users could be imprisoned without trial for being suspected of
using an anonymity system, then 
an opt-in strategy might be the most ethical one.

\paragraph{Unethical even with opt in}
Yet another ethical position is that deploying conscripted
anonymity is not ethical under any circumstances, even 
when using an opt-in mechanism.
One argument supporting this position is that the risks
of being conscripted into an anonymity system might not
be clear to a novice Web user, even after the user is presented
with a description of the conscripted anonymity system.
If the user does not have the technical understanding to make
an informed decision about the risks of opting in to the
system, then it might not {\em ever} be ethical to offer
users the option to participate in \name.

\vspace{\baselineskip}

\Name is compatible with both ``opt out'' and ``opt in'' policies, so the
decision of which policy to use can be left to the
organization deploying such a system.
Different policies will be appropriate in different
societal contexts.
In contexts where the risks
to conscripted users are high, using \name
may not be appropriate at all.

\section{Implementation and Evaluation}
\label{sec:impl}

\abbr{}{
This section presents evaluation results for our
proof-of-concept implementation of a \name JavaScript client.

\subsection{\Name Client}}
To evaluate the performance of
\name, we have implemented
the JavaScript client applications for two 
anonymity systems: a timed cascade mix-net
and a verifiable client/server DC-net.
Our prototypes perform all of 
the cryptographic operations that a full-featured
JavaScript client would perform but they do not yet
produce messages that are wire-compatible with the 
underlying anonymity systems.

To be useful, the \name JavaScript client must be able
to produce at least one dummy 
message before the user browses away to another page.
One recent survey of Web usage finds that the median 
time spent on a Web page is 11 seconds~\cite{kumar10characterization},
so the JavaScript application should generate at 
least one dummy message every few seconds.

We tested each prototype on four platforms: a 
modern Linux workstation (Ubuntu 13.04), 
a Mac laptop (Mac OS 10.6), an iPhone 4 (iOS 6),
and a Motorola Android phone (Android 2.2).
We tested the JavaScript client on each device
using the Chrome 26, Firefox 21, Safari 5, and Opera Mobile 12
browsers, where available.
\abbr{}{
We used ``Web workers'' (essentially JavaScript threads) 
to prevent the browser GUI from freezing
while it executes long-running cryptographic operations.
}

\paragraph{Mix-net and Remailer}
To simulate the casual user workload for a mix-net or
anonymous remailer, we created a JavaScript application
that encrypts a 256-byte message with five layers of 
RSA-2048 public key encryption using the
OpenPGP.js JavaScript library~\cite{openpgpjs}.
\abbr{}{
}
Table~\ref{tab:timing} presents results for the
mix-net client.
Our results demonstrate that even a CPU-limited
mobile device can generate a mix-net dummy message
in fewer than 10 seconds, which is
less time than the median time spent on a Web page 
(11 seconds, as explained above).
\abbr{
The extended version of this paper sketches a method
for distributing the mix servers' public
keys to \name users.
}{

To make \name work with a real mix-net
or remailer, there must be some way of distributing the
public keys of the mix servers to the users (both casual
and savvy) of the system.
One possible distribution method would be to have a set
of directory authorities (as in Tor~\cite{dingledine04tor})
who collective maintain the list of public keys.
The savvy users would have a signature verification public key
for each of the directory authorities hard-coded into their
\name browser plug-in.
Whenever the list of mix servers changes (e.g., because a new
mix server comes online), all of the directory authorities
sign the list of keys and then distribute the signed list 
to the Web servers.

When the Web server serves the JavaScript client application
to a user, the server includes the latest list of public keys
in a block of static text on the page.
The casual user's \name JavaScript application would read in the
public keys and generate a dummy message.
A savvy user's browser plug-in first checks the served JavaScript
(as described in Section~\ref{sec:attack:javascript}), then 
reads in the list of public keys from the page, verifies the signatures
of each of the entropy authorities on the list, 
and then uses the list of public keys to encrypt its message.
}

\paragraph{Verifiable DC-net}
To simulate a casual user's workload when using
a more computationally intensive anonymity system, we
implemented the client functionality for a verifiable
DC-net~\cite{corrigangibbs13verif,golle04dining} 
using the 
Stanford JavaScript cryptography library~\cite{sjcl-ecc}.
\abbr{}{
Generating a dummy message for the verifiable DC-net requires
the client to perform elliptic curve operations and 
to generate a non-interactive zero-knowledge proof.
}
Our evaluation uses 
NIST's P-256 elliptic curve group~\cite{nist09fips186} 
and requires the client to generate a 32-byte dummy message.
The performance results (Table~\ref{tab:timing}) for the
verifiable DC-net suggest that the \name JavaScript application 
is arguably practical on faster machines, 
since both the workstation
and laptop were able to generate a dummy message in 
less than four seconds each.
Performance on the CPU-limited 
iPhone is less impressive---generating
a single message took at least 30 seconds.

\abbr{}{
In a real-world deployment of \name,
the Web browser might serve the JavaScript application 
only to machines that the server expects will be able
to generate dummy messages quickly.
For example, the server could avoid sending the application 
to browsers running on the iOS or Android 
operating systems, as determined by the HTTP {\tt User-Agent} header. 
}

\newcommand*\raiseup[1]{%
\begingroup
\setbox0\hbox{\tiny\strut #1}%
\leavevmode
\raise\dimexpr \ht\strutbox - \ht0\box0
\endgroup
}

\begin{table}[t]
\centering
\begin{tabular*}{0.48\textwidth}[t]{l r | r r}
& & Mix-net& DC-net\\
\hline
{\bf Workstation} &Chrome&81&156\\
\raiseup{\footnotesize \em Intel W3565 3.20 GHz}&Firefox&73&1,781\\
  \hline
{\bf Laptop} & Chrome&133&231\\
\raiseup{\footnotesize \em Intel Core 2 Duo 2.53 GHz} &Firefox&171&3,062\\
&Safari&669&3,338\\
  \hline
{\bf Apple iPhone 4} &Chrome&9,009&62,973\\
\raiseup{\footnotesize \em Apple A4 (speed unknown)}&Safari&7,280&32,972\\
  \hline
{\bf Motorola Milestone} &Opera&$\dagger$&63,504\\
\raiseup{\footnotesize \em ARM Cortex-A8 600 MHz}&\multicolumn{1}{c}{}&\multicolumn{2}{l}{}\\
\multicolumn{4}{p{0.47\textwidth}}{$\dagger$ %
{\footnotesize Opera Mobile does not support the {\tt getRandomValues}
API required by OpenPGP.js.}}
\end{tabular*}
\caption{Time (milliseconds) to generate 
a \name ``dummy'' user message in JavaScript using different
hardware/browser combinations.}
\label{tab:timing}
\end{table}

\abbr{}{
\subsection{Power Consumption on Mobile Devices}
\label{sec:power}

\begin{figure}[t]
\centering
\includegraphics[width=0.45\textwidth]{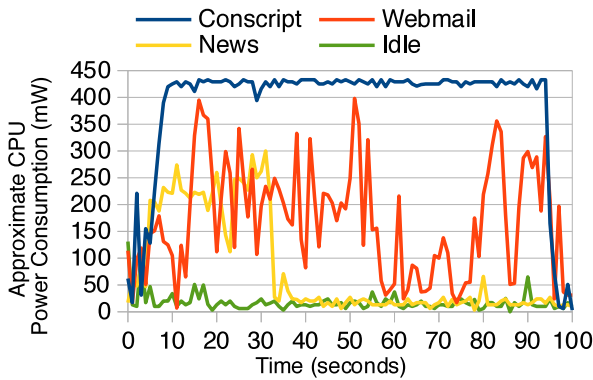}
\caption{Approximate CPU power consumption on Android
  phone during Web browsing and anonymity conscription.} 
\label{fig:eval-power}
\end{figure}

Battery life is an important limitation on mobile devices.  
If a Web page
serving a \name JavaScript client consistently drained a phone's battery,
mobile users might begin to opt out of participation in \name or
even avoid visiting the Web site entirely.

We evaluate the impact of \name on 
mobile devices by measuring the power consumption due 
to CPU usage on the Android phone when the phone's user
was idle, was browsing a news Web site (\url{nytimes.com}),
was interacting with a Webmail service (\url{gmail.com}),
and was running the \name JavaScript client
for the verifiable DC-net anonymity system described above. 
The trace for the \name client measures the power consumed
when the device generates a DC-net ciphertext corresponding to a
32-byte dummy message.
Our experiment used the PowerTutor 1.5~\cite{powertutor} Android
application to estimate power usage in each scenario.

Figure~\ref{fig:eval-power} plots a trace of the power 
usage for each of the four browsing scenarios
and Table~\ref{tab:power} summarizes the 
average power usage and energy consumption for
each scenario.
Generating a verifiable DC-net dummy message using 
the \name JavaScript client uses
just over $2.25\times$ more energy than the phone
used while browsing the Webmail application for the
same length of time.
Generating the dummy message consumed nearly $5\times$ more
energy than the phone consumed while visiting the news Web site
for the same length of time.
However, if a Web user visits a \name-enabled
Web site only a few times per day,
the user might not notice the two-fold increase
in power consumption during these sessions.

To reduce the energy consumption of the \name JavaScript
client, the Web site serving the application could
reduce the participation rate for mobile devices,
as described in 
\abbr{the extended version of this paper}{Section~\ref{sec:arch:rate}}.
If Web sites serving the JavaScript client are particularly
sensitive to consuming more power on users' devices,
the Web sites detect mobile users 
(e.g., using the HTTP {\tt User-Agent} header)
and could avoid serving the JavaScript application 
to these users entirely.

\begin{table}
\centering
\begin{tabular}{l | r r}
\multicolumn{1}{l}{}& \multicolumn{1}{l}{Average} & \multicolumn{1}{l}{Energy} \\
& \multicolumn{1}{l}{Power (mW)}& \multicolumn{1}{l}{Consumed (mWh)}\\
    \hline
Idle &17.7 & 0.52\\
News&74.5&2.19\\
Webmail&162.8&4.79\\
Conscript&376.7&10.82\\
\end{tabular}
\caption{Average power usage and total
  energy consumption on Android device
  for different browsing scenarios.}
\label{tab:power}
\end{table}
}

\section{Related Work}
\label{sec:rel}
The FlashProxy system~\cite{fifield12evading}, which was one of the
inspirations for this work, uses a JavaScript
application to coerce Web users into serving as
bridges into the Tor anonymity network~\cite{dingledine04tor}.
Every additional Web browser that runs the FlashProxy application 
increases the {\em access} to the Tor network in regions where
Tor relays are blocked. 
In contrast, every additional Web browser that runs
the \name JavaScript application increases the {\em anonymity}
available to users of the anonymity system.

Anonymity systems have used {\em dummy messages} in the
past to deter traffic analysis 
attacks~\cite{berthold02dummy,moeller00mixmaster}.
However, these systems use dummy messages only {\em inside of}
the anonymity network and they do not have
``dummy users'' send messages into the system
to increase the effective number of total users of
the system.

Bauer~\cite{bauer03new} describes a system for using
a specially crafted banner ads served to 
an unwitting user's Web browser to create a covert channel
between two servers in a mix network.
Bauer considers only {\em passive} adversaries, whereas
we consider {\em active} adversaries
that also can monitor all network traffic.
Since \name considers a stronger threat model, we address 
a number of security issues in Section~\ref{sec:attack} that
Bauer's work did not consider.

AdLeaks~\cite{roth13secure}---a system design 
published independently 
while this work was in preparation---%
uses JavaScript served by online advertising networks
to conscript users into participation in 
an anonymity system.
Unlike \name, which is general and compatible
with a number of existing anonymity system, 
AdLeaks conscripts users only
into AdLeaks' own anonymity system.
In addition, the AdLeaks anonymity system is {\em not} designed to prevent
active attacks by dishonest participants in the system,
whereas \name can protect against such attacks.
\abbr{}{
In the AdLeaks system,
each conscripted user submits an encryption
of zero to the anonymity system while savvy users submit
encryptions of plaintext messages.
If two savvy users submit a message in the same 
transmission ``slot,'' a collision results and the
messages are undecipherable.
AdLeaks uses error-correcting codes to allow communication
in the presence of accidental collisions.
However, a malicious user could induce collisions in {\em every} AdLeaks
transmission slot, which would halt communication entirely.
\Name, when used with a mix-net or verifiable DC-net (as evaluated in
Section~\ref{sec:impl}), avoids the need for error-correcting codes and is
not vulnerable to these anonymous denial-of-service attacks.
}

\abbr{}{
CryptoCat~\cite{kobeissi13cryptocat} is an encrypted chat system---%
not an anonymity system---which
performs cryptographic operations in a Web browser plug-in, much
as we propose for conscripted anonymity's savvy users.
}

\section{Conclusion}
\label{sec:concl}

\abbr{}{
We have described how to use JavaScript to make it easier
for casual Web users to participate in strong anonymity systems.
These {\em conscripted anonymity sets} can increase
the anonymity available to users of 
mix-nets, verifiable shuffle systems, 
and remailers.
}
We have presented \name, general architecture for
conscripted anonymity, we discuss a number of attacks
against \name (and possible defenses), 
we consider ethical issues of deploying such a system, and we
implement and evaluate a proof-of-concept prototype
on a variety of devices.
\Name can increase the user base
of formally analyzable, but unpopular, anonymity systems, 
which allows the few security-sensitive users of these systems
to hide amongst a larger group of casual Internet users.

\subsection*{Acknowledgements}
We would like to thank David Fifield, David Wolinsky,
and the anonymous reviewers
for their helpful comments on the draft.
This material is based upon work supported by the Defense Advanced Research
Agency (DARPA) and SPAWAR Systems Center Pacific, Contract No. 
N66001-11-C-4018.


{\small
\bibliographystyle{abbrv}
\bibliography{net,os,sec,soc}
}

\end{document}